\documentclass[final,5p,times,twocolumn]{elsarticle}

\usepackage{lineno,hyperref}
\usepackage{amsmath}
\usepackage{lipsum}
\usepackage{stfloats}
\modulolinenumbers[5]
\biboptions{sort&compress}
\usepackage{xtab}

\journal{International Journal of Heat and Mass Transfer}

    \makeatletter
\def\ps@pprintTitle{%
	\let\@oddhead\@empty
	\let\@evenhead\@empty
	\def\@oddfoot{\reset@font\hfil\thepage\hfil}
	\let\@evenfoot\@oddfoot
}
\makeatother

\begin{document}

\begin{frontmatter}

\title{Transient heat transfer of superfluid $^4$He in nonhomogeneous geometries - Part I: Second sound, rarefaction, and thermal layer}

\author[address1]{Shiran Bao}
\author[address1,address2]{Wei Guo\corref{corresponding}}
\cortext[corresponding]{Corresponding author}
\ead{wguo@magnet.fsu.edu}

\address[address1]{National High Magnetic Field Laboratory, 1800 East Paul Dirac Drive. Tallahassee, FL 32310, USA}
\address[address2]{Mechanical Engineering Department, Florida State University, Tallahassee, FL 32310, USA}

\begin{abstract}
Transient heat transfer in superfluid $^4$He (He II) is a complex process that involves the interplay of the unique counterflow heat-transfer mode, the emission of second-sound waves, and the creation of quantized vortices. Many past researches focused on homogeneous heat transfer of He II in a uniform channel driven by a planar heater. In this paper, we report our systematic study of He II transient heat transfer in nonhomogeneous geometries that are pertinent to emergent applications. By solving the He II two-fluid equation of motion coupled with the Vinen's equation for vortex-density evolution, we examine and compare the characteristics of transient heat transfer from planar, cylindrical, and spherical heaters in He II. Our results show that as the heater turns on, an outgoing second-sound pulse emerges, in which the vortex density grows rapidly. These vortices attenuate the second sound and result in a heated He II layer in front of the heater, i.e., the thermal layer. In the planar case where the vortices are created throughout the space, the second-sound pulse is continuously attenuated, leading to a strong thermal layer that diffusively spreads following the heat pulse. On the contrary, in the cylindrical and the spherical heater cases, vortices are created mainly in a thin thermal layer near the heater surface. As the heat pulse ends, a rarefaction tail develops following the second-sound pulse, in which the temperature drops. This rarefaction tail can promptly suppress the thermal layer and take away all the thermal energy deposited in it. The effects of the heater size, heat flux, pulse duration, and temperature on the thermal-layer dynamics are discussed. We also show how the peak heat flux for the onset of boiling in He II can be studied in our model.
\end{abstract}


\end{frontmatter}


\section{Introduction}
Saturated liquid $^4$He transits to the superfluid phase (known as He II) below about 2.17 K \cite{Tilley-book}. In He II, two miscible fluid components co-exist: an inviscid and zero-entropy superfluid component (i.e., the condensate) and a viscous normal-fluid component (i.e., the collection of thermal excitations). This two-fluid system possesses many fascinating thermal and mechanical properties \cite{landau-1987}. For instance, He II supports two distinct sound modes: an ordinary pressure-density wave (i.e., the first sound) where the two fluids move in phase, and a temperature-entropy wave (i.e., the second sound) where the two fluids move oppositely. Furthermore, heat transfer in He II is via an extremely effective counterflow mode instead of convection \cite{vansciver-2012}: the normal fluid carries the heat away from a source at a velocity $v_n$=$q/\rho sT$, where $q$ is the heat flux, and $\rho$ and $s$ are the He II density and specific entropy, respectively; while the superfluid moves in the opposite direction at a velocity $v_s$=$-v_n\rho_n/\rho_s$ to balance the mass flow (here $\rho_n$ and $\rho_s$ are the densities of the respective fluid components). When the relative velocity of the two fluids exceeds a small critical value, a chaotic tangle of quantized vortex lines can be spontaneously created in the superfluid, each carrying a quantized circulation $\kappa\simeq10^{-3}$ cm$^2$/s around its angstrom-sized core~\cite{donnelly-1991}. A mutual friction force between the two fluids appears due to the scattering of the thermal excitations off the quantized vortices \cite{vinen-1957_Proc.R.Soc.Lond.A_II}. This mutual friction can profoundly affect the heat transfer and turbulence characteristics in both fluids \cite{marakov-2015_Phys.Rev.B, Gao-2016-JETP, Gao-2017-PRB, Gao-2017-JLTP, Gao-2018-PRB, Bao-2018-PRB,Mastracci-2018-PRF}.

Due to its low temperature and extraordinary heat-transfer capability, He II has been widely utilized in scientific and engineering applications such as for cooling superconducting particle accelerator cavities, superconducting magnets, and satellites \cite{vansciver-2012}. Many of these applications involve transient heat transfer in He II, a process that is complicated due to the interplay of counterflow, second-sound emission, and vortex nucleation. There have been extensive experimental and numerical studies of one-dimensional (1D) transient heat transfer of He II in a uniform channel driven by a planar heater, due to the simplicity of this geometry~\cite{fiszdon-1989_J.LowTemp.Phys.,fiszdon-1990_J.FluidMech.,shimazaki-1995_Cryogenics,hilton-2005_J.LowTemp.Phys.,zhang-2006_Int.J.HeatMassTransf.}. These studies have revealed that the transient heating from the heater generates a second-sound pulse that propagates in He II. A counterflow establishes in the pulse, which produces tangled quantized vortices. These vortices then attenuate the second-sound pulse, converting the energy carried by the pulse to the internal energy of He II. This heated region in front of the heater is termed as the thermal layer. When the heat flux is relatively high, the continuous attenuation eventually curtails the second-sound pulse to a limiting profile \cite{shimazaki-1995_Cryogenics}, and the heat produced by the heater largely gets deposited in the thermal layer which gradually diffuses along the channel following the second-sound pulse.

It has been recognized that the heat transfer of He II in nonhomogeneous geometries can exhibit new features. For instance, Fiszdon \emph{et al.} conducted transient heat transfer experiments in He II using cylindrical heaters~\cite{fiszdon-1990_J.FluidMech.}. They found that a rarefaction tail of the second-sound pulse can develop, which exhibits a drop in temperature. The thermal layer in this geometry can be significantly suppressed as compared to that in the planar geometry. These observations were examined and reproduced in numerical simulations by Kondaurova, \emph{et al.}~\cite{kondaurova-2008_J.LowTemp.Phys.,kondaurova-2017_J.LowTemp.Phys.,kondaurova-2020_LowTemp.Phys.}. Nevertheless, there lacks a systematic characterization of the thermal-layer dynamics and how the heat energy is divided between the thermal layer and the propagating second sound. Producing this knowledge could benefit applications pertinent to cylinder shaped systems cooled by He II, such as superconducting transmission lines and magnet coils~\cite{maksoud-2010_IEEETrans.Appl.Supercond,xavier-2019_IEEETrans.Appl.Supercond}. An emergent effort in developing hot-wire anemometry for studying quantum turbulence in He II~\cite{duri-2015_Rev.Sci.Instrum} has further strengthened this need.

Besides the cylindrical geometry, transient heat transfer of He II in spherical geometry is also relevant to practical applications. In particular, it has been known that superconducting accelerator cavities cooled by He II can quench due to transient heating from tiny surface defects~\cite{conway-2017_Supercond.Sci.Technol.}. Locating these surface hot spots for subsequent defect removal is the key for improving the cavity performance. Our team has recently developed an innovative molecular tagging technique for locating surface hot spots via tracking thin lines of He$^*_2$ molecular tracers~\cite{bao-2019_Phys.Rev.Applied, bao-2020_Int.J.HeatMassTransf.}. These tracers move with the normal fluid~\cite{Guo-2009-PRL,Guo-2010-PRL,Gao-2015-RSI,Guo-2014-PNAS}, and therefore the transient radial heat transfer from a hot spot can lead to line deformations that contain accurate information about the spot location. In order to extract this information, it is critical to understand how the heat energy is partitioned between the thermal layer and the second-sound pulse~\cite{bao-2019_Phys.Rev.Applied}. However, despite some limited studies on steady-state vortex distribution near a spherical heater \cite{inui-2020_Phys.Rev.B,Varga-2019-JLTP}, the transient behaviors of the thermal layer and its interaction with the second sound in this geometry have remained largely unexplored~\cite{kondaurova-2008_J.LowTemp.Phys.}.

In this paper, we present a numerical study of transient heat transfer in all three (i.e., planar, cylindrical, and spherical) geometries in He II. Our goal is to examine and compare the heat-transfer characteristics in these geometries so that a better understanding of the energy partition and thermal-layer dynamics can be achieved. The paper is organized as follows. In Sec.~\ref{sec:model}, we introduce our model, which is based on the governing equations of the two-fluid system and the Vinen's equation for vortex-density evolution. In Sec.~\ref{sec:validation}, we validate our model by comparing the simulation result of transient heat transfer in planar and cylindrical geometries with the experimental measurements by Fiszdon \emph{et al.}~\cite{fiszdon-1990_J.FluidMech.}. The systematic study of the heat transfer in all geometries is discussed in Sec.~\ref{sec:result}. We first present in Sec.~\ref{sec:wave} the calculated spatial profiles of the second-sound wave, the vortex-line density, and the thermal energy in the three geometries under the same heating conditions. This comparison clearly shows the unique features of He II heat transfer in nonhomogeneous geometries. We then discuss the thermal-layer dynamics in the cylindrical and the spherical geometries and the effects of various heat-pulse parameters in Sec.~\ref{sec:tl}. In Sec.~\ref{sec:peak}, we illustrate how our model can also be used to determine the peak heat flux for the onset of boiling in He II in different geometries. A summary is included in Sec.~\ref{sec:conclusion}.

\begin{figure*}[htb]
\center
\includegraphics[width=0.9\linewidth]{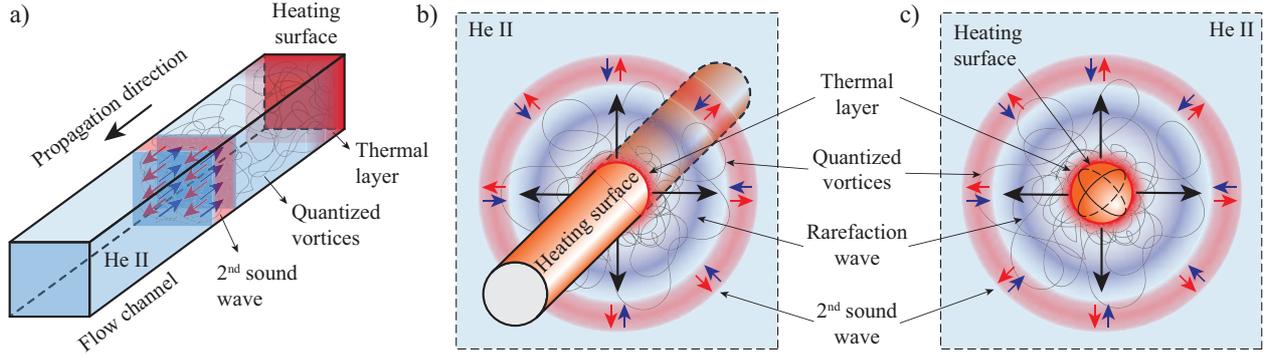}
\caption{Schematic diagrams showing the transient heat transfer in He II from (a) a planar heater, (b) a cylindrical heater, and (c) a spherical heater.} \label{Fig1}
\end{figure*}

\section{\label{sec:model} Numerical model}
To study the flow field and the heat transfer in He II, we adopt the Hall-Vinen-Bekarevich-Khalatnikov (HVBK) model~\cite{Nemirovskii-1995-RMP}, which is based on the conservation equations for the fluid mass, momentum, and entropy, as listed below:
\begin{equation}
	\frac{\partial \rho}{\partial t}+\mathbf{\nabla} \cdot(\rho \mathbf{v})=0
	\label{eq:mass}
\end{equation}
\begin{equation}
	\frac{\partial(\rho s)}{\partial t}+\mathbf{\nabla} \cdot(\rho s\mathbf{v_n})=\frac{\mathbf{F_{ns}}\cdot\mathbf{v_{ns}}}{T}
	\label{eq:energy}
\end{equation}
\begin{equation}
	\frac{\partial \mathbf{v_s}}{\partial t}+\mathbf{v_s} \cdot \mathbf{\nabla} \mathbf{v_s}+\mathbf{\nabla} \mu=\frac{\mathbf{F_{ns}}}{\rho _s}
	\label{eq:vs}
\end{equation}
\begin{equation}
	\frac{\partial (\rho \mathbf{v})}{\partial t}+\mathbf{\nabla} (\rho_sv_s^2+\rho_nv_n^2)+\mathbf{\nabla} p=\eta_n\mathbf{\nabla} v_n^2
	\label{eq:v}
\end{equation}
The definitions of the relevant parameters are provided in the Nomenclature table. In the above equations, $\rho \mathbf{v}=\rho_s \mathbf{v_s}+\rho_n \mathbf{v_n}$ represents the total momentum density. The Gorter-Mellink mutual friction $\mathbf{F_{ns}}$ per unit fluid volume depends on the vortex-line density $L$ and the relative velocity $\mathbf{v_{ns}}=\mathbf{v_{n}}-\mathbf{v_{s}}$ between the two fluids as~\cite{vinen-1957_Proc.R.Soc.Lond.A_II,vinen-1957_Proc.R.Soc.Lond.A_III}:
\begin{equation}
	\mathbf{F_{ns}}=\frac{\kappa}{3}\frac{\rho_s\rho_n}{\rho}B_LL\mathbf{v_{ns}}
	\label{eq:Fns}
\end{equation}
where $B_L$ is a known temperature-dependent mutual friction coefficient~\cite{Donnelly-1998-JPCRD}. The chemical potential $\mu(P,T,v_{ns})$ of He II includes a correction due to the counterflow velocity $v_{ns}$, as proposed by Landau~\cite{landau-1987}:
\begin{equation}
\mu(P,T,v_{ns})=\mu(P,T)-\frac{1}{2}\frac{\rho_n}{\rho}v_{ns}^2
\label{eq:landau_mu}
\end{equation}
This HVBK model represents a coarse-grained description of the two-fluid hydrodynamics, since the action of individual vortices on the normal fluid \cite{Yui-2020-PRL, Mastracci-2019-PRF} is smoothed out. When the vortex-line density is relatively high, this model has been shown to describe non-isothermal flows in He II very well even in nonhomogeneous geometries \cite{Sergeev-2019-EPL, zhang-2006_Phys.Rev.B}.

To provide a closure to the above HVBK model, we adopt a modified version of the Vinen's phenomenological equation to determine the temporal and spatial variations of the vortex-line density $L(\mathbf{r},t)$~\cite{vinen-1957_Proc.R.Soc.Lond.A_II,vinen-1957_Proc.R.Soc.Lond.A_III}:
\begin{equation}
	\frac{\partial L}{\partial t}+\mathbf{\nabla} \cdot (\mathbf{v_L}L)=\alpha_V|v_{ns}|L^{3/2}-\beta_VL^2+\gamma_V|v_{ns}|^{5/2}
	\label{eq:l}
\end{equation}
where $\alpha_V$, $\beta_V$ and $\gamma_V$ are temperature-dependent phenomenological coefficients introduced by Vinen~\cite{vinen-1957_Proc.R.Soc.Lond.A_II}. The term $\mathbf{\nabla}\cdot(\mathbf{v_L}L)$ accounts for the drifting of the vortices~\cite{schwarz-1988_Phys.Rev.B, nemirovskii-2019_LowTemp.Phys.}, where the vortex mean velocity $\mathbf{v_L}$ is taken to be the local superfluid velocity $\mathbf{v_s}$, as originally proposed by Vinen~\cite{vinen-1957_Proc.R.Soc.Lond.A_II,vinen-1957_Proc.R.Soc.Lond.A_III} and later utilized by many others~\cite{wang-1987_Phys.Rev.B,kondaurova-2014_Phys.Rev.B}. The first two terms on the right-hand side of Eq.~\ref{eq:l} respectively account for the generation and the decay of the vortices, and the third source term serves to trigger the initial growth of the line density~\cite{vinen-1957_Proc.R.Soc.Lond.A_II}.

If one ignores the vortices and linearizes Eqs.~(\ref{eq:mass})-(\ref{eq:v}) assuming small-amplitude wave-form variations of the entropy and the counterflow velocity, it is straightforward to derive a temperature-entropy wave mode (i.e., the second sound)~\cite{landau-1987}. A transient heating from a heater surface then generates a second-sound pulse in He II whose amplitude $\Delta T$ is determined by the heat flux. When this amplitude is relatively high, the second-sound speed $c_{2}$ can be written as $c_2=c_{20}[1+\varepsilon(T)\Delta T]$, where $c_{20}$ is the speed in the zero-amplitude limit and the nonlinear coefficient $\varepsilon(T)$ takes the form~\cite{landau-1987}:
\begin{equation}
\varepsilon(T)=\frac{\partial}{\partial T}\mathrm{ln}\left(\frac{c_{20}^{3}C_p}{T}\right)
\label{eq:c2}
\end{equation}
At $T<1.88$ K where $\varepsilon(T)$ is positive, the second-sound wave with a higher amplitude travels faster. Therefore, a front shock can appear at the leading edge of the second-sound pulse at sufficiently large $\Delta T$. At $T>1.88$ K where $\varepsilon(T)$ is negative, a rear shock can form at the tail of a second-sound pulse. This physical picture gets complicated when vortices are present, which can attenuate and distort the second-sound pulse profile.

Here we consider the transient heat transfer from planar, cylindrical, and spherical heaters in He II based on all the coupled governing equations (i.e., Eqs.~\ref{eq:mass}-\ref{eq:l}), as shown schematically in Fig.~\ref{Fig1}. For simplicity, we ignore small-scale turbulent fluctuations and assume 1D flow in all three geometries, i.e., 1D flow perpendicular to the heater in the planar case and along the radial direction in the cylindrical and the spherical cases. For a rectangular heat pulse with a surface heat flux $q_h$ and a duration $\Delta t$, we set the boundary conditions at the heater surface to be $v_n=q_h/\rho sT$ for the normal fluid and $v_s=-v_n\rho_n/\rho_s$ for the superfluid during $0<t<\Delta t$ and $v_n=v_s=0$ at $t>\Delta t$. All the thermodynamic properties of He II are calculated using the Hepak dynamic library~\cite{hepak-2005}. The values of the coefficients $\alpha_V$ and $\beta_V$
as recommended by Kondaurova \emph{et al.} are used in Eq.~\ref{eq:l}, which appear to produce simulation results in good agreement with experimental observations~\cite{kondaurova-2017_J.LowTemp.Phys.}. We then evolve the governing equations using the MacCormack's predictor-corrector scheme, which is accurate to the second order in time and space~\cite{fletcher-2003}. A flux-corrected transport approach is also adopted to suppress the numerical oscillations due to the discontinuity at the shock front~\cite{fletcher-2003}. We have tested various spatial steps $\Delta r$ and time steps $\Delta t_s$ and found that the calculated results converged well when $\Delta r<2\times10^{-5}$ m and $\Delta t_s<2\times10^{-8}$ s. In order to balance the result fidelity and the computational cost, $\Delta r=10^{-5}$~m and $\Delta t_s=10^{-8}$ s are used in all the reported simulations.

\section{\label{sec:validation}Model Validation}
\begin{figure}[h]
	\centering
	\includegraphics[width=1\linewidth]{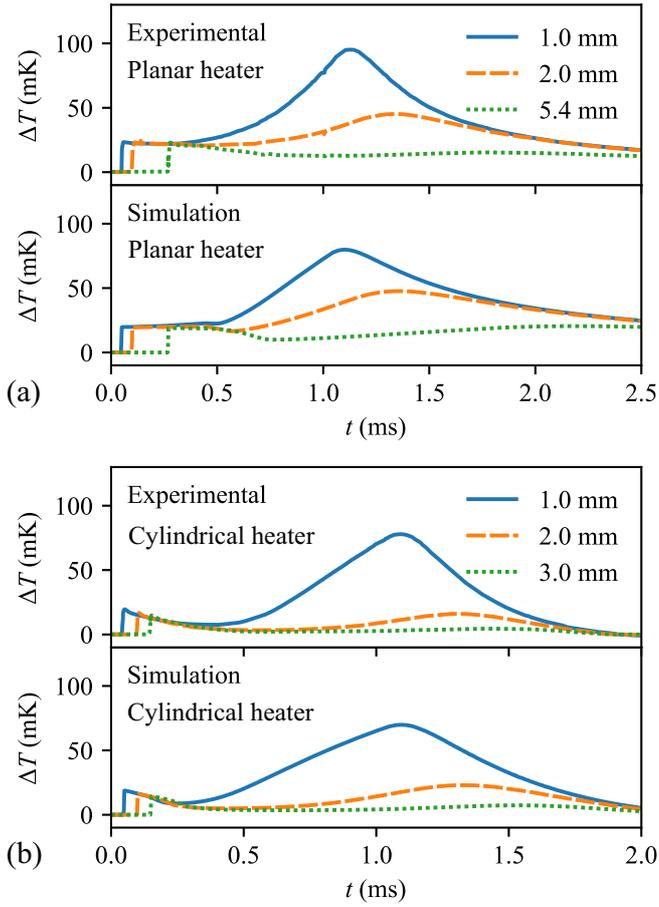}
	\caption{Experimental and simulated temporal profiles of the temperature increment $\Delta T=T-T_\infty$ at (a) 1, 2, and 5.4 mm from the surface of a planar heater with $q_h$=5 W/cm$^2$, $\Delta t$=1 ms , and a repetition rate of 0.2 Hz; and (b) at 1, 2, and 3 mm from the surface of a cylindrical heater with $q_h$=6 W/cm$^2$, $\Delta t$=1 ms, $r_h$=2.5 mm, and a repetition rate of 2 Hz. The bath temperature is $T_\infty=1.4$ K.} \label{fig:vali1D}
\end{figure}

For model validation purpose, we first performed numerical simulation on transient heat transfer of He II under the same conditions as in the experiments conducted by Fizdon \emph{et al.}~\cite{fiszdon-1990_J.FluidMech.}. These authors examined the transient heat transfer from both a planar heater and a cylindrical heater (radius $r_h=2.5$ mm) immersed in He II at 1.4 K. For the planar heater case, they used heat pulses with a duration $\Delta t=1$ ms at a repetition rate of 0.2 Hz, and the heat flux on the heater surface was $q_h=5$ W/cm$^2$. They measured the time variations of the He II temperature at distances $r=1$ mm, 2 mm, and 5.4 mm from the heater surface using a movable superconducting bolometer. In their experiment, the vortices generated by one heat pulse did not have enough time to decay when the next heat pulse arrived. Therefore, the initial vortex-line density $L_0$ seen by a given heat pulse was relatively high, which is often treated as a tuning parameter in past numerical works~\cite{fiszdon-1989_J.LowTemp.Phys.,fiszdon-1990_J.FluidMech.,kondaurova-2008_J.LowTemp.Phys.}. In our calculation, we set $L_0=8\times10^5$ cm$^{-2}$ to achieve the best match with the experimental observations. Fig.~\ref{fig:vali1D}(a) shows the measured temperature profiles together with our simulation results. Since $\varepsilon(T)>0$ at 1.4 K, the temperature at each location first spikes up sharply upon the arrival of the shock front of the second-sound pulse. A gradual temperature overshot is then observed, which is due to the spreading of the thermal layer as we shall discuss in details in Sec.~\ref{sec:result}. All these observations are well reproduced in our simulations.

In the cylindrical heater case, the measured and the simulated temperature variations at distances $r-r_h=1$ mm, 2 mm, and 3 mm from the heater surface are shown in Fig.~\ref{fig:vali1D}(b). In this case, heat pulses with a duration $\Delta t=1$ ms and a surface heat flux $q_h=6$ W/cm$^2$ were applied at 2 Hz repetition rate. Due to the nonhomogeneous geometry, a radial dependance of the initial line density $L_0(r)=L_h(r_h/r)^2$ as recommended by Kondaurova \emph{et al.}~\cite{kondaurova-2008_J.LowTemp.Phys.} was adopted in our calculation, where the line density at the heater surface $L_h$ was set to 8$\times10^6$ cm$^{-2}$ due to the increased repetition rate. Again, all the key features of the observed temperature curves are reproduced. This excellent agreement between the experimental measurements and our simulation results has thereby validated the fidelity of our model calculation.

\section{\label{sec:result}Simulation results and discussion}
In this section, we first present the simulation results to compare the key features associated with the transient heat transfer in different heater geometries. We then examine the time evolution of the thermal layer in the cylindrical and the spherical heater cases under various heating conditions. Since our focus is the heat transfer following a single heat pulse, a small initial vortex-line density $L_0=10^2$ cm$^{-2}$ is assumed in the calculations. This $L_0$ is comparable to the typical density of remnant vortices pinned to He II container walls~\cite{Awschalom-1984-PRL}. Indeed, it has been shown that in relatively high flux counterflow, the simulated temperature profile in He II is nearly independent of $L_0$ when $L_0$ is smaller than about $10^5$ cm$^{-2}$ due to the source term in Eq.~\ref{eq:l}~\cite{kondaurova-2020_LowTemp.Phys.}. To avoid the complication of possible boiling in He II near the heater surface, we have also assumed that the heater is placed at a 1-meter depth below the He II free surface in all the cases. We will discuss in the last subsection how this hydrostatic head pressure ensures the helium to be always in the He II state during the transient heat transfer. This discussion also provides a foundation for our future study of the peak heat flux for the onset of boiling in He II.

\subsection{\label{sec:wave} Transient heat transfer characteristics in different heater geometries}
\begin{figure*}[htb]
	\centering
	\includegraphics[width=1\linewidth]{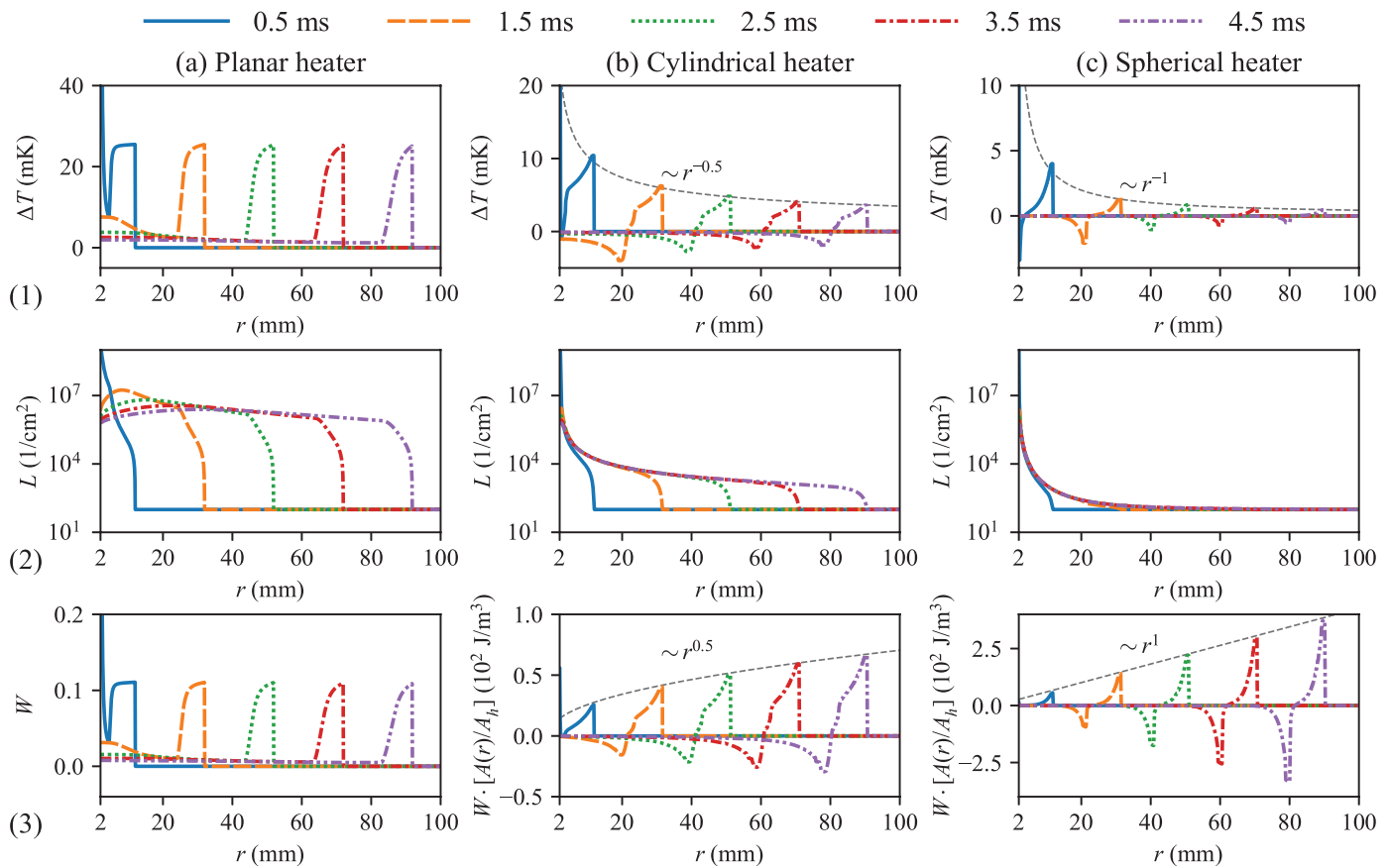}
	\caption{Profiles of (1) temperature increment $\Delta T$, (2) vortex-line density $L$, and (3) thermal energy density $W$ compensated by the ratio of the cross section area $A(r)$ to the heater surface area $A_h$ in (a) planar, (b) cylindrical, and (c) spherical geometries at 1.78 K. In all cases, $q_h$=23 W/cm$^2$, $\Delta t$=0.5 ms and $r_h$=2 mm.} \label{fig:TLq}
\end{figure*}
To compare the heat transfer characteristics in different heater geometries, we show the simulated spatial profiles of the temperature increment $\Delta T=T-T_\infty$, the vortex-line density $L$, and the thermal energy density $W=\rho C_p\Delta T$ at various time $t$ in Fig.~\ref{fig:TLq}. In this calculation, we set the He II bath temperature to $T_\infty=1.78$ K. A heat pulse with a surface flux $q_h=23$ W/cm$^2$ and a fixed duration $\Delta t=0.5$ ms is applied to the heater at $t=0$ in all three cases. The radii of both the cylindrical and the spherical heaters are set to $r_h=2$ mm. In what follows, we discuss the main features of the transient heat transfer.

\textbf{1) Second-sound pulse:} As shown in Fig.~\ref{fig:TLq}(1a-1c), a second-sound pulse with positive $\Delta T$ emerges when the heater turns on, which carries the heat energy and propagates away from the heater surface at the known second-sound speed (i.e., $c_2=19.6$ m/s at 1.78 K \cite{Donnelly-1998-JPCRD}) in all three cases. Inside the pulse profile, a counterflow establishes where the normal-fluid velocity is determined by the thermal energy flux as $v_n=c_2W/\rho sT=(c_2C_p/s)\cdot\Delta T/T$. This counterflow leads to a rapid generation of the quantized vortices. In the planar heater case, the second-sound pulse gradually evolves from a rectangular profile near the heater to a front-shock profile due to the combined effects of the vortex attenuation and the positive $\varepsilon(T)$. In the cylindrical and the spherical heater cases, as the second-sound pulse propagates outward, the cross-section area of the pulse $A(r)$ increases as $r$ and $r^2$, respectively. In regions where the vortex density is low and hence the mutual friction is negligible, the kinetic energy of each fluid component is nearly conserved. For the normal fluid, this means that $v_n$ in the second-sound pulse must drop as $1/\sqrt{r}$ in the cylindrical geometry and as $1/r$ in the spherical geometry. Since $\Delta T$ is proportional to $v_n$, it also drops in a similar fashion as the pulse propagates, which is clearly seen in Fig.~\ref{fig:TLq}(1b-1c).

\textbf{2) Quantized vortices:} The vortices are created as a consequence of the counterflow in the second-sound pulse. In the planar heater case, the thermal energy flux $W$ in the pulse remains high as the pulse propagates. Therefore, a dense tangle of vortices are created in the entire space traversed by the second-sound pulse (see Fig.~\ref{fig:TLq}(1a)), which continuously attenuate the pulse. In the cylindrical and the spherical heater cases, since the thermal energy flux drops with $r$ due to the diverging geometries, the line density $L$ is high (i.e., greater than 10$^5$ cm$^{-2}$) only in a thin layer of He II near the heater surface (see Fig.~\ref{fig:TLq}(1b-1c)). Outside this region, the second-sound pulse experiences negligible attenuation.

\textbf{3) Rarefaction tail:} A peculiar feature of the temperature profile in the cylindrical and the spherical geometries, as compared to the planar case, is the appearance of a tail region with negative $\Delta T$ following the positive second-sound pulse (see Fig.~\ref{fig:TLq}(1b-1c)). This negative $\Delta T$ tail, which emerges after the heater is switched off, is known as the rarefaction wave~\cite{efimov-1998_LowTemp.Phys.,fiszdon-1989_Phys.FluidsA,iznankin-1983_Sov.Phys.JETP}. The underlying physics can be understood as follows. The total thermal energy carried by the second-sound pulse can be evaluated as $Q_s=\int_{\Delta R}W(r)A(r)dr$, where $\Delta R\simeq c_2\Delta t$ is the thickness of the pulse. Since $W(r)A(r)$ is expected to increase as $\sqrt{r}$ in the cylindrical geometry and as $r$ in spherical geometry (confirmed in our simulation, i.e., see Fig.~\ref{fig:TLq}(3b-3c)), $Q_s$ increases as the pulse propagates. To supply this ever-growing thermal energy carried by the second-sound pulse, there must be a flow of the internal energy from the tail region towards the pulse front, which thereby leads to the formation of the negative $\Delta T$ rarefaction tail. If we integrate $Q_s$ over both the positive pulse and the rarefaction tail, the total thermal energy carried by them always equals the input heat energy, which fulfills the energy-conservation law. 

\begin{figure*}[htb]
	\centering
	\includegraphics[width=1\linewidth]{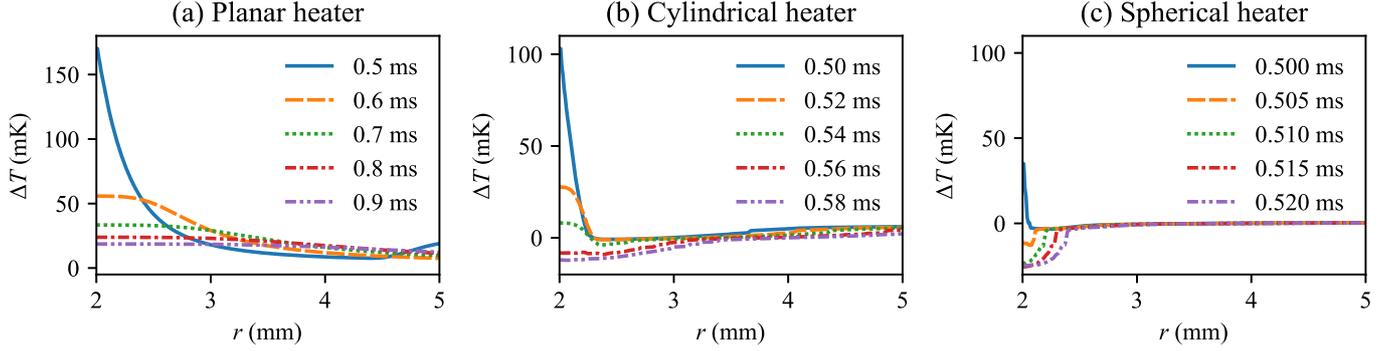}
	\caption{Profiles of the temperature increment $\Delta T$ near the heater surface in (a) planar, (b) cylindrical, and (c) spherical geometries at 1.78 K. In all cases, $q_h=23$ W/cm$^2$, $\Delta t=0.5$ ms and $r_h=2$ mm.} \label{fig:TTL}
\end{figure*}

\textbf{4) Thermal layer:} Near the heater surface where the vortex-line density $L$ is high, the interaction between the vortices and the second-sound pulse effectively converts the thermal energy carried by the pulse to locally deposited heat, resulting in a heated layer of He II, i.e., the thermal layer. To see this layer clearly, we plot the $\Delta T$ profile near the heater in all three cases in Fig.~\ref{fig:TTL}. As the heat pulse ends, $\Delta T$ on the heater surface reaches the highest value. In the planar heater case, $\Delta T(r_h)=170$ mK on the heater surface, which is about 7 times the $\Delta T$ in the second-sound pulse. The heat content in this thermal layer diffusively spreads out \cite{vansciver-2012}. On the contrary, in the cylindrical and the spherical heater cases, the temperature buildup in the thermal layer is much weaker. Indeed, both the layer thickness and the maximum $\Delta T$ in the spherical geometry are insignificant. Another important feature of the thermal-layer dynamics in the two nonhomogeneous geometries is that this layer dies out rapidly before it has time to undergo diffusive spreading. This prompt suppression is due to the same mechanism for the formation of the rarefaction tail: the internal energy in these nonhomogeneous geometries is actively transferred towards the second-sound pulse front in order to supply the ever-growing thermal energy carried by the pulse. The depletion of the deposited heat in the thermal layer occurs simultaneously with the formation of the rarefaction tail.

\begin{figure}[htb]
	\centering
	\includegraphics[width=0.9\linewidth]{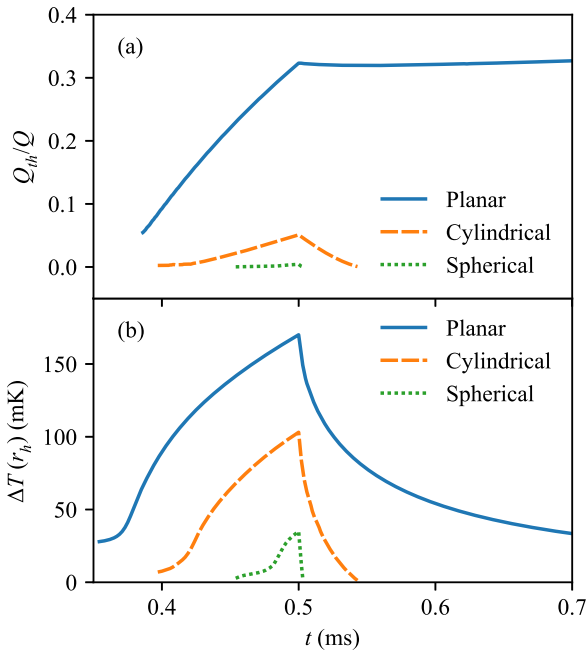}
	\caption{Evolution of (a) the fraction of the heat energy $Q_{th}/Q$ deposited in the thermal layer; and (b) the temperature increment $\Delta T (r_h)$ at the heater surface at 1.78 K. In all three cases, $q_h=23$ W/cm$^2$, $\Delta t=0.5$ ms, and $r_h=2$ mm.} \label{fig:Geo}
\end{figure}
\textbf{5) Heat energy partition:} The partition of the heat energy between the thermal layer and the second-sound pulse as well as how this partition varies with time is of practical significance. To examine this partition, we calculate the heat energy in the thermal layer $Q_{th}$ by integrating $W$ from the heater surface $r_h$ to the boundary $r_b$ of the thermal layer as $Q_{th}=\int^{r_b}_{r_h}W(r)A(r)dr$. This boundary $r_b$ in the planar heater case is set to be the minimum temperature location between the heater surface and the second-sound pulse. For the cylindrical and the spherical heater cases, $r_b$ is set to the location where $\Delta T$ drops to zero. In Fig.~\ref{fig:Geo}, we show the calculated time evolution of the ratio of $Q_{th}$ to the input heat $Q=q_hA_ht$ (where $q_h=23$ W/cm$^2$ at $0\leq t\leq \Delta t$) as well as the $\Delta T (r_h)$ on the heater surface for all three cases. The fraction $Q_{th}/Q$ in the planar heater case increases to about 35\% by the end of the heat pulse, and it slowly increases even after the heater is turned off since the second-sound pulse keeps producing vortices and experiencing attenuation. Following the heat pulse, the thermal layer spreads out and therefore $\Delta T (r_h)$ decreases. In the cylindrical and the spherical heater cases, $Q_{th}/Q$ only reaches 5\% and 0.4\%, respectively, with $\Delta T (r_h)=100$ mK and 26 mK by the end of the heat pulse. Furthermore, $Q_{th}$ quickly drops to zero due to the aforementioned mechanism that occurs in nonhomogeneous geometries. In the end, all the input heat energy is completely carried away by the second-sound pulse and the rarefaction tail.

\subsection{\label{sec:tl} Effects of heating conditions and bath temperature on the thermal-layer dynamics}
In this subsection, we present more detailed studies on the thermal-layer dynamics in the two nonhomogeneous geometries under various heating conditions and bath temperatures. 

\begin{figure}[htb]
	\centering
	\includegraphics[width=0.95\linewidth]{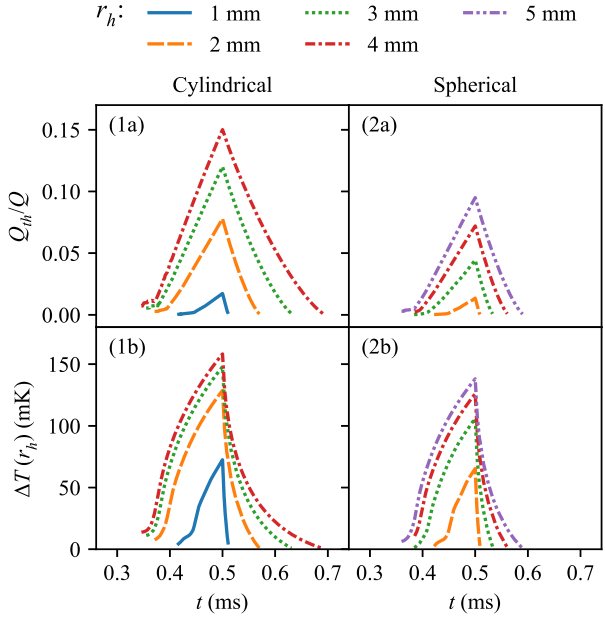}
	\caption{Effect of the heater radius $r_h$ on the evolution of (a) $Q_{th}/Q$ and (b) $\Delta T (r_h)$ in (1) the cylindrical and (2) the spherical heater geometries. In both cases, $T_\infty=1.78$ K, $q_h=24$ W/cm$^2$, and $\Delta t=0.5$ ms.} \label{fig:r0}
\end{figure}

\begin{figure}[!]
	\centering
	\includegraphics[width=0.95\linewidth]{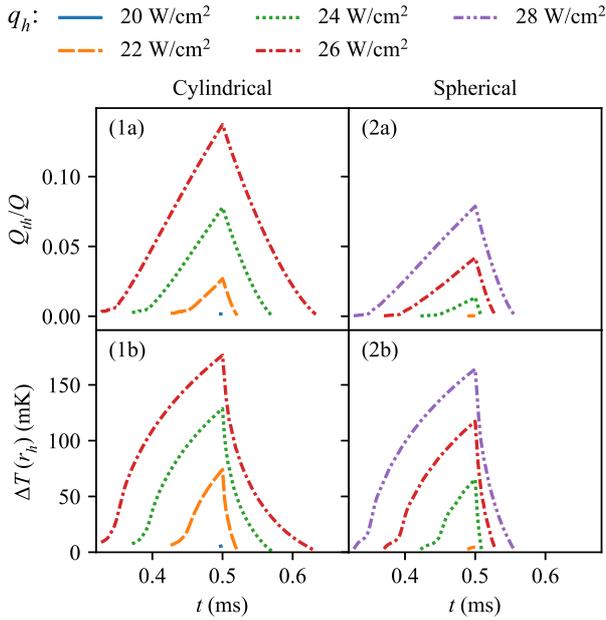}
	\caption{Effect of the surface heat flux $q_h$ on the evolution of (a) $Q_{th}/Q$ and (b) $\Delta T (r_h)$ in (1) the cylindrical and (2) the spherical heater geometries. In both cases, $T_\infty=1.78$ K, $\Delta t=0.5$ ms, and $r_h=2$ mm.} \label{fig:Q}
\end{figure}

\begin{figure}[htb]
	\centering
	\includegraphics[width=0.95\linewidth]{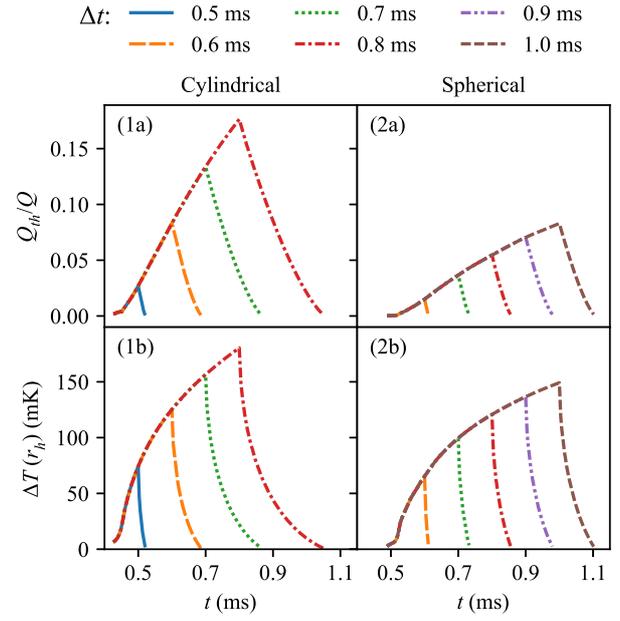}
	\caption{Effect of the pulse duration $\Delta t$ on the evolution of (a) $Q_{th}/Q$ and (b) $\Delta T (r_h)$ in (1) the cylindrical and (2) the spherical heater geometries. In both cases, $T_\infty=1.78$ K, $q_h=24$ W/cm$^2$, and $r_h=2$ mm.} \label{fig:th}
\end{figure}

\textbf{1) Effects of heating conditions:} First, we vary the heater radius $r_h$ in the range of 1 mm to 5 mm while keeping the same surface heat flux $q_h=24$ W/cm$^2$, pulse duration $\Delta t=0.5$ ms, and bath temperature $T_\infty=1.78$ K. The results are shown in Fig.~\ref{fig:r0}. It is clear that both the deposited heat energy $Q_{th}$ and the surface temperature increment $\Delta T (r_h)$ increase with the heater size. This is not surprising, since the heater surface appears flatter to the adjacent He II at larger $r_h$. Therefore, the thermal-layer dynamics is expected to evolve towards that in the planar geometry as $r_h$ increases. Fig.~\ref{fig:Q} shows the results with a varying surface heat flux $q_h$ in the range of 20 to 28 W/cm$^2$ at fixed pulse duration $\Delta t=0.5$ ms, heater radius $r_h=2$ mm, and bath temperature $T_\infty=1.78$ K. As the surface heat flux $q_h$ increases, the thermal layer starts to grow earlier and can reach a higher $\Delta T(r_h)$ with more deposited energy $Q_{th}$. Interestingly, we see that when $q_h$ is lower than a threshold $q^{(c)}_h$, i.e., about 20 W/cm$^2$ in the cylindrical geometry and about 22 W/cm$^2$ in the spherical geometry, the thermal layer does not grow at all. In the planar case, $q^{(c)}_h$ is significantly lower, i.e., about 15 W/cm$^2$ in our simulation, in agreement with the reported values~\cite{vansciver-2012, shimazaki-1995_Cryogenics}. Although the exact value of $q^{(c)}_h$ depends on other heating parameters and the bath temperature, it is always higher in nonhomogeneous geometries under given heating conditions. Finally, we show the effect of the pulse duration $\Delta t$ in Fig.~\ref{fig:th}, where $\Delta t$ is varied from 0.5 ms to 1.0 ms at fixed $q_h=22$ W/cm$^2$, $r_h=2$ mm, and $T_\infty=1.78$ K. The deposited heat energy $Q_{th}$ increases almost linearly with $\Delta t$ in both geometries.


\begin{figure}[!]
	\centering
	\includegraphics[width=0.95\linewidth]{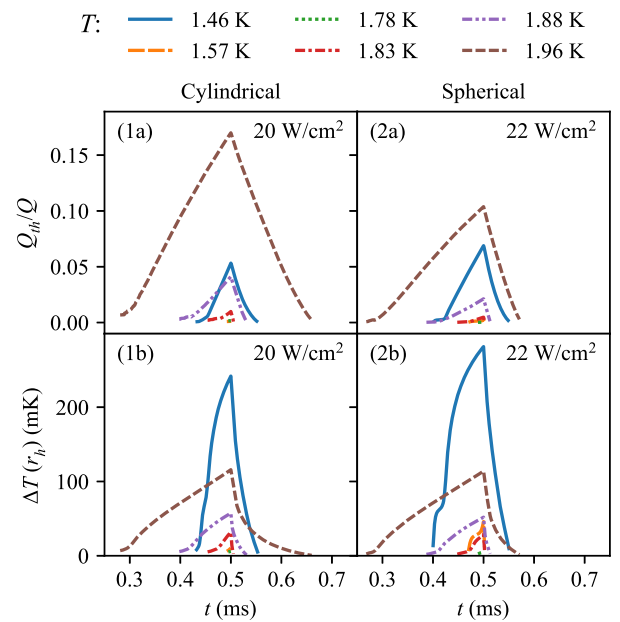}
	\caption{Effect of the bath temperature $T_\infty$ on the evolution of (a) $Q_{th}/Q$ and (b) $\Delta T (r_h)$ in (1) the cylindrical heater geometry with $q_h=20$ W/cm$^2$ and (2) the spherical heater geometry with $q_h=22$ W/cm$^2$. In both cases, $\Delta t=0.5$ ms and $r_h=2$ mm.} \label{fig:T}
\end{figure}

\begin{figure}[htb]
	\centering
	\includegraphics[width=1\linewidth]{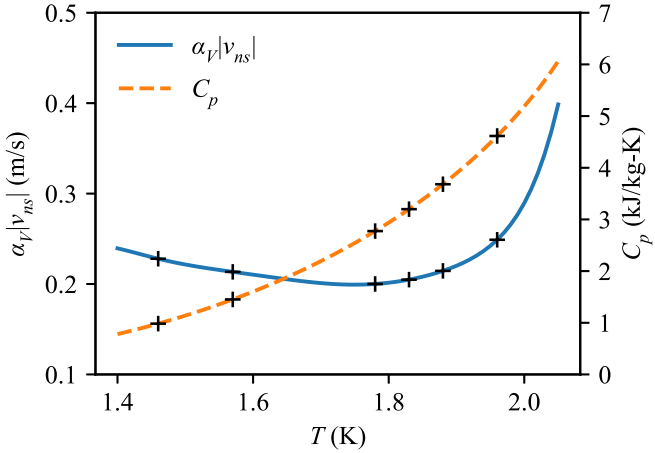}
	\caption{The temperature dependance of the coefficient $\alpha_V|v_{ns}|$ and the He II heat capacity $C_p$. The crosses mark the temperatures examined in Fig.~\ref{fig:T}.} \label{fig:Teffect}
\end{figure}

\textbf{2) Effect of bath temperature:} The effect of the bath temperature $T_\infty$ on the thermal-layer dynamics is more complicated, since nearly all the parameters in the governing equations are temperature dependent but their variations with $T_\infty$ can be quite different. Fig.~\ref{fig:T} shows our calculation results when $T_\infty$ is varied from 1.46 K to 1.96 K while all other heating conditions remain fixed, i.e., $r_h=2$ mm, $\Delta t=0.5$ ms, and $q_h=20$ W/cm$^2$ in the cylindrical geometry and $q_h=22$ W/cm$^2$ in the spherical geometry. As $T_\infty$ increases from 1.46 K, the maximum heat energy deposited in the thermal layer first decreases and becomes negligible when $T_\infty$ is in a range of roughly 1.6 K to 1.8 K. Then, as $T_\infty$ further increases, the deposited energy starts to rise. This non-monotonic behavior can be qualitatively understood as caused by the temperature dependance of $\alpha_V|v_{ns}|$ in the vortex generation term in Eq.~(\ref{eq:l}). In Fig.~\ref{fig:Teffect}, we plot $\alpha_V|v_{ns}|$ as a function of temperature, where $v_{ns}$=$q/\rho_s sT$ is evaluated at the surface heat flux $q_h=22$ W/cm$^2$. It is clear that $\alpha_V|v_{ns}|$ exhibits a non-monotonic temperature dependance and reaches a minimum value at around 1.75 K. Since the generation term in Eq.~(\ref{eq:l}) largely controls the rate of vortex production, for a fixed pulse duration $\Delta t$, the vortex-line density in the thermal layer is low at small $\alpha_V|v_{ns}|$. Consequently, the attenuation to the second-sound pulse is weak, which limits the heat energy deposited in the thermal layer. We also note that at the lowest temperature that we have examined, i.e., $T_\infty=1.46$ K, the highest $\Delta T (r_h)$ on the heater surface is achieved despite the fact that only up to 5\% of the heat energy is deposited. This pronounced temperature change is essentially caused by the small heat capacity of He II at low temperatures. As shown in Fig.~\ref{fig:Teffect}, the heat capacity drops rapidly with decreasing the temperature. At low $T_\infty$, even a small energy deposition in the thermal layer can therefore result in a large temperature increment.

\subsection{\label{sec:peak} Boiling in He II during transient heat transfer}
An important parameter in He II heat transfer applications is the so-called peak heat flux $q_c$~\cite{vansciver-2012}. This $q_c$ denote the heat flux from the heater surface above which boiling occurs in the helium. There have been extensive studies of $q_c$ in the planar geometry, and existing correlations can reasonably predict the values of $q_c$ at various temperatures and heating conditions~\cite{vansciver-2012}. However, the studies on $q_c$ in nonhomogeneous geometries are very limited. Here, we discuss how our model may allow us to systematically evaluate $q_c$ for He II transient heat transfer in different geometries.

\begin{figure}[t]
	\centering
	\includegraphics[width=1\linewidth]{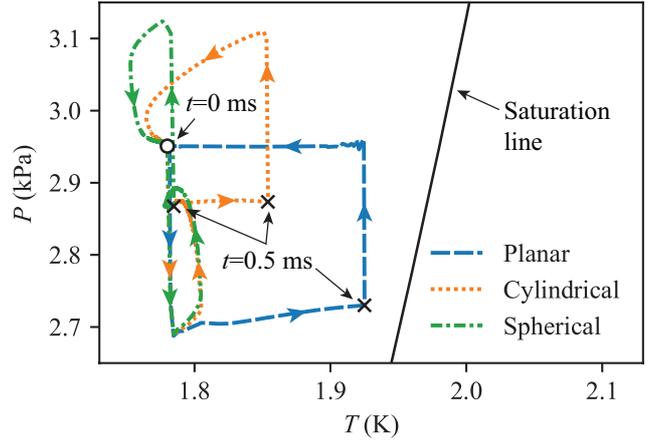}
	\caption{The evolution of the He II state near the heater surface. In all cases, $T$=1.78 K, $q_h$=23 W/cm$^2$, $\Delta t$=0.5 ms, and $r_h$=2 mm. The circle and the crosses mark the start and the end of the heat pulse, respectively.} \label{fig:PT}
\end{figure}

Note that in all the cases we have studied, the highest temperature in He II is always achieved near the heater surface. Therefore, we just need to monitor the state of the He II adjacent to the heater during a transient heat transfer. If at any instance, the state in the $P$-$T$ phase diagram traverses the saturation line to the vapor phase or the He I phase, boiling is deem to occur. Fig.~\ref{fig:PT} shows representative $P$-$T$ curves of the He II near the heater surface at $T_\infty=1.78$ K, $q_h=23$ W/cm$^2$, $\Delta t=0.5$ ms, and $r_h=2$ mm in three heater geometries. As the heater turns on, a sudden drop in the local pressure is seen in all three cases. Then, the state curve moves horizontally towards the saturation line in the planar heater case, while in the two nonhomogeneous geometries the pressure rises a bit before the state curve moves horizontally. Under the specified heating conditions, the horizontal move in the spherical geometry is negligible. After the heat pulse ends, the pressure spikes up in all the cases and the $P$-$T$ curves then evolve back to the start point. The complex evolution pathes in the nonhomogeneous geometries are intimately related to the rarefaction physics. From this example calculation, one can see clearly that the He II state approaches the saturation line furthermost in the planar heater case. If we increases the surface heat flux $q_h$, the state curve in the planar geometry will touch the saturation line first, which allows us to determine the $q_c$ in this geometry. The $q_c$ in the other two geometries can be determined in a similar fashion as we further increase $q_h$. It is clear that under the same heating conditions, $q_c$ is the lowest in the planar geometry and is the highest in the spherical geometry. We may then vary parameters such as the hydrostatic pressure head, the pulse duration $\Delta t$, and the heater radius $r_h$ to study their effects on $q_c$. The relevant details will be presented in a future publication.

\section{\label{sec:conclusion} Conclusion}
We have conducted numerical simulations of transient heat transfer in He II by solving the HVBK equations of motion for the two-fluid system coupled with the Vinen's equation for the evolution of quantized vortices. The characteristics of transient heat transfer from planar, cylindrical, and spherical heaters are systematically examined. Compared to the planar heater case, the heat transfer in the nonhomogeneous geometries exhibits some distinct key features. These include: 1) a rapid drop of the vortex-line density away from the heater surface; 2) the formation of a thin thermal layer near the heater beyond which the second-sound pulse experiences negligible attenuation; 3) the emergence of a rarefaction tail with a negative temperature increment following the second-sound pulse; and 4) a prompt suppression of the thermal layer upon the completion of the heat pulse such that all the input heat can be completely carried away by the outgoing second-sound pulse and the rarefaction tail. We have also examined the effects of various heating parameters and the He II bath temperature on the evolution of the thermal layer. Our result shows that the thermal layer diminishes more quickly with a smaller heater size, a lower surface heat flux, or a shorter pulse duration. When the heating conditions are fixed, the buildup of the thermal layer exhibits a non-monotonic dependence on the bath temperature. To invoke the topic in the Part II of our series of reports, we have also illustrated how our model will allow us to study the peak heat flux for the onset of boiling in He II during a transient heat transfer. These studies should provide us a solid foundation towards the development of a comprehensive understanding of He II transient heat transfer in different geometries.

\appendix
\section*{Acknowledgement}
The authors would like to thank Dr. S. W. Van Sciver for valuable discussions. This research is supported by the U.S. Department of Energy under Grant No. DE-SC0020113 and was conducted at the National High Magnetic Field Laboratory at Florida State University, which is supported through the National Science Foundation Cooperative Agreement No. DMR-1644779 and the state of Florida.

\section*{Nomenclature}

\xentrystretch{-0.15}
\renewcommand{\arraystretch}{1}

\tabletail{%
	\hline
}

\tablehead{%
	\hline
}

\begin{xtabular}[H]{p{1cm}p{5.6cm}p{1.1cm}}
	\hline	
	Variable 			& Description \rule{0pt}{3ex}							& Unit 						\\	
	\hline
	$A$					& Area													& m$^2$						\\
	$B_L$				& Mutual friction coefficient							&							\\
	$c_2$				& Second sound velocity									& m/s						\\
	$c_{20}$			& Second sound velocity at zero amplitude				& m/s						\\
	$C_p$				& Heat capacity at constant pressure					& J/(kg$\cdot$K) 			\\
	$F_{ns}$			& Mutual friction force									& N/m$^3$		 			\\
	$h$					& Planck constant										& J$\cdot s$				\\
	$L$       			& Vortex line density  								 	& 1/m$^2$					\\
	$m$                 & Mass of the helium atom								& kg						\\
	$p$					& Pressure												& Pa						\\
	$q$					& Heat flux												& W/m$^2$					\\
	$Q$					& Input energy											& J							\\
	$Q_s$				& Energy carried by the second-sound pulse				& J							\\
	$Q_{th}$			& Energy in the thermal layer							& J							\\
	$r$					& Coordinate											& m							\\
	$r_b$				& Boundary of the thermal layer							& mm						\\
	$r_h$				& Heater radius											& mm						\\
	$s$					& Entropy												& J/(kg$\cdot$K) 			\\
	$t$					& Time													& s							\\
	$T$					& Temperature											& K							\\
	$T_\infty$			& Helium bath temperature								& K							\\
	$v$					& Velocity												& m/s						\\
	$v_L$				& Vortex mean velocity									& m/s						\\
	$v_{ns}$			& Counterflow velocity, $v_n-v_s$						& m/s						\\
	$W$					& Thermal energy density								& J/m$^3$  					\\
	\\
	$Greeks$        	                                                      	                    		\\
	$\alpha_V$        	& Coefficient in Vinen's equation                      	&                   		\\
	$\beta_V$        	& Coefficient in Vinen's equation                      	& m$^2$/s           		\\
	$\gamma_V$        	& Coefficient in Vinen's equation                      	& s$^{1.5}$/m$^{4.5}$ 		\\
	$\Delta r$			& Spatial step											& m 						\\
	$\Delta R$			& Thickness of second-sound pulse						& m 						\\
	$\Delta t$			& Heating duration										& s 						\\
	$\Delta t_s$		& Time step												& s 						\\
	$\Delta T$			& Temperature difference, $T-T_\infty$					& K							\\
	$\varepsilon (T)$   & Nonlinear coefficient for the second sound			& 1/K						\\
	$\eta_n$			& Viscosity												& Pa$\cdot$s				\\
	$\kappa$			& Quantum of superfluid circulation, $h/m$				& m$^2$/s					\\
	$\mu$				& Chemical energy										& J/kg						\\
	$\rho$				& Density												& kg/m$^3$					\\ 	
	\\
	$Subscripts$    	                                                       	                    		\\
	$h$          		& Heater surface						               	&                   		\\
	$n$          		& Normal fluid component                               	&                   		\\
	$s$          		& Superfluid component                               	&                   		\\
	$0$					& Initial state											&							\\
	\hline
\end{xtabular}

\bibliography{References}

\end{document}